# Enhancing Consumer User Experience, Education and Brand Awareness through Musical Advergames

Forouzan FARZINNEJAD[a], Hadi KHEZRIAN, Mohsen KASIRI, Nilufar BAGHAEI[b*]
[a]Department of Computer Engineering, Sharif University of Technology, Tehran, Iran
[b]School of Natural and Computational Sciences, Massey University, Auckland, New Zealand
farzinnejad@ce.sharif.edu; Hadikhezrian@yahoo.com; mohsenkasiri181373@gmail.com;
N.Baghaei@massey.ac.nz

**Abstract:** The effectiveness of traditional communication techniques has declined in recent years, and marketing specialists are looking at more creative ways to engage consumers. A lot of attention has recently been paid to advergames, which are seen as an attractive new marketing tool to increase consumer engagement and education of brand awareness. The term advergame refers to games that combine brand advertising with gaming to promote business products. In this paper, we study the impact of music on consumer uptake of advergames (n=197). Our results show that music in advergames is an important feature because of its capability to attract audience, enhance their user experience and increase brand awareness.

## 1. Introduction

Advertising is a marketing communication that employs an openly sponsored, nonpersonal message to promote or sell a product, service, or idea. Sponsors of advertising are typically businesses. Advertisements deliver the message in creative ways to reach as many audiences as possible [Ho & Lam, 2019]. "We breathe nitrogen, oxygen, and advertising every day!" says French advertising critic Robert Grange. Advertising as a cultural phenomenon has a significant impact on all aspects of our social life [Qin, 2019]. In today's advertising industry, one of the ways to increase interest in advertising is the field of entertainment programming. Digital Games as an attractive mass media have become an international marketing communication tool for advertisers [Bellman et al, 2014].

Advergaming combines a promotional message, company logo or other brand information with online games or video games. This has created an appealing and engaging platform that is becoming increasingly popular with the widespread use of social media. Advergames are regarded as innovative tools for building a customer relationship, which are both suitable for existing products and business startups [Waiguny et al, 2011]. Research findings indicate that advergames are a promising form of advertising, particularly for young people [Bellman et al, 2014]. However, there are a significant number of gamers of all ages. Also, the fact that 44% of all gamers are women shows that the dominance of males in the field of digital gaming is fading [Lee & Cho 2017]. A strategy followed by the majority of media planners these days is the consumer choice of media that enhances brand recall through entertainment. Therefore, advertising managers should also incorporate games into the design of advergames to ensure that their implementation positively influences consumers' memory [Vashisht, & Sreejesh 2015]. The Advergaming technique helps increase brand awareness and remembrance, which persuades consumers to make a purchase.

## 1.1 Immersion and Persuasive

Persuasive technology is defined as technology that is designed to change attitudes or behaviors of the users through persuasion and social influence [Fogg, 2002]. Such technologies are used in sales and marketing, education, politics, military training, public health, lifestyle changes, and management, and may potentially be used in any area of human-human or human-computer interaction. Most self-identified persuasive technology research focuses on interactive, computational technologies, including desktop computers, web services, mobile devices [Reddy et al, 2017], video games [Oinas-Kukkonen & Harjumaa, 2008] and more recently, virtual reality [Samarasinghe, Baghaei & Stemmet, 2020]. In spite of variations in design and appearance, an important common element in many successful computer games is the ability to attract and engage players. This is called the experience of immersion, a word widely used by gamers and critics [Jennett et al, 2008] – meaning one's entire mind is concentrated on the game as part of the user experience, aka deep mental involvement.

## 1.2 The Role of Music and Sound in Video Games & Advertising

Sound is an essential element in video games [Cunningham et al, 2006]. Central to UX in digital gaming is deliberate treatment of sound and music to affect the player's feedback and rewards. In video games, audio UX is shaped by sound and audio cues. The role of audio is to facilitate interaction with the virtual gaming world and trigger interactive feedback [Nacke et al, 2010]. Because sounds relate to usability, they provide information that directly responds to or demands the player's performance. In other cases, sounds have a more informative function in which they direct the player's orientation or help the player identify different situations and situations [Jørgensen, 2008]. Background music, which is a collective term used to refer to music and sound effects, has long been considered as an inherent element of modern video games [Zhang & Fu 2015] and has proven to be an essential component of any popular video game [Areni & Kim, 1993, Politis et al, 2016]. Given that game production has become more complicated than in the past and carries high costs (up to millions of dollars), sound engineers and composers have assumed great importance.

Advertising makes use of music as a driving component in commercials to enrich the original message. The concept of environmental processing indicates that environmental cues like music can result in a good attitude towards advertising and towards the brand [Morris & Boone, 1998]. An environmental cue such as music has the most impact on brand attitudes in an advertising environment. The significant association between music and emotions has indicated that exciting music enhances emotional arousal in individuals manifested through skin reactions and increased heart pulsation, which are both physiological symptoms for emotional response. Music influences brand cognition and purchase intention in shaping and facilitates purchase intention [Morris & Boone, 1998]. Research has confirmed the profound influence of music on arousal, mood, and emotion. In fact, our brain responds to more melodies than language. Music evokes emotions, and emotions make decisions such as buying or not buying a product. So, it can be used strategically to increase sales. In general, music appears to increase cognitive load, which is an effective way to get people thinking. Research conducted by the Nielsen Institute has confirmed that music helps to connect brands with their customers on an emotional level [Nielsen, 2019].

Experts believe that genre influences consumption habits. In a study [Areni & Kim, 1993], a wine shop began playing classical music instead of top billboard songs. As a result, it sold more expensive bottles of wine. Studying music in restaurants has also confirmed the correlation between classical music and rising costs. Consumers, therefore, are likely to have heard classical music associated with complexity and expensive tastes and have inadvertently altered their behavior to suit the environment.

In advertising, memorability is important - more effective advertising stays in memory. Although consumers are expected to remember the message of an advertisement, this will directly lead to shopping only if accompanied by emotional involvement. In fact, the best advertisements are those which are both informative and emotionally powerful. When an advertisement uses music, the audience strives to identify and remember the original message, which in turn improves business sentiment and customers buy from their favorite brand.

Our project examines the effectiveness of providing music on consumers' uptake of advergames and enhancing their user experience. In this paper, we outline the design, implementation, findings and plan for the future work.

## 2. Implementation & Evaluation Study

We conducted a study at the International Electronic, Computer & E-Commerce Exhibition, Tehran, Iran. It is the greatest commercial event in Iran's Market of Electronics and Computer Products and services to advertise the first mobile brand, the world's first and largest mobile and high-speed network operator in the country. The target audience of the games were people of different age and gender.

The Megabox [Boks, 2020] machine is an advergame display device that has an industrial monitor that is vertically positioned. On the screen of the monitor is a touch module so that the user can interact with it. The coordinates (x, y) of a point on the touch module are sent depending on what point the user touches. The device also includes a speaker for audio playback. All devices are connected to the Internet so that they can text the bonus code to the user or print it on paper using a printer embedded in the device. To identify and prevent participants from playing a game more than once, they were asked to enter their phone numbers into megabox.

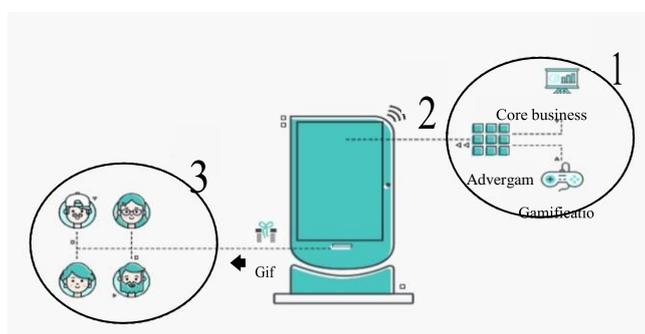

*Figure 1.* Megabox mechanism

This game we chose for this study was adopted from Piano tiles [Wikipedia, 2020]. The participants first selected one of the activities and services related to the brand's in-app icons, including charging purchases, Internet pack purchases, etc. By selecting the charging icon, the service provided features such as inserting a charging card, charging a purchase, charging purchase history and custom charging. Instead of black tiles in the original piano tiles game, the services were presented in circle tiles. When each services circle tile is tapped by the user, it will make a sound of cooperate music we intend to promote instead of piano tile music. If the player taps on a white tile, the player will lose the game and be signaled by an off-tune note. Similar to Piano Tile, the game had three levels of speed, and after touching a certain number of circles, the game speed increased. It should be noted that touching each of these circles increased the player's score by one. As the participant's score increased, more valuable prizes remained and fewer valuable prizes were eliminated. The purpose of this scenario was to convey the intended advertising message (brand features of the app) by advergame.

A chance mechanism was also designed to reward participants. Sixteen bonus boxes appeared with more valuable prizes depending on the player's score in the game. The participants selected one of the remaining bonus boxes, and the prize would appear. The participants were able to receive their prize by presenting the brand name printed by Megabox to the brand booth. For the study, we used two Megabox machines, with two different versions of the game (Piano advergame with and without background music). The numbers of females and males and the participants' age are as follow:

Table 1: *The number of females and males*

|  | Females | Males | Total |
|---|---|---|---|
| Piano tiles with background music | 46% | 54% | 124 |

| | Piano tiles without background music | 52.1% | 47.9% | 73 |
|---|---|---|---|---|

During the gameplay, the services, features, new features of that company's application, product or brand were passed on to the target audience, depending on the advertising part of an advergame, and nature and need of the brand. At the end of a session, according to the game's aspect in the advergame and the need for a reward, the player was rewarded by the brand, proportion to their score.

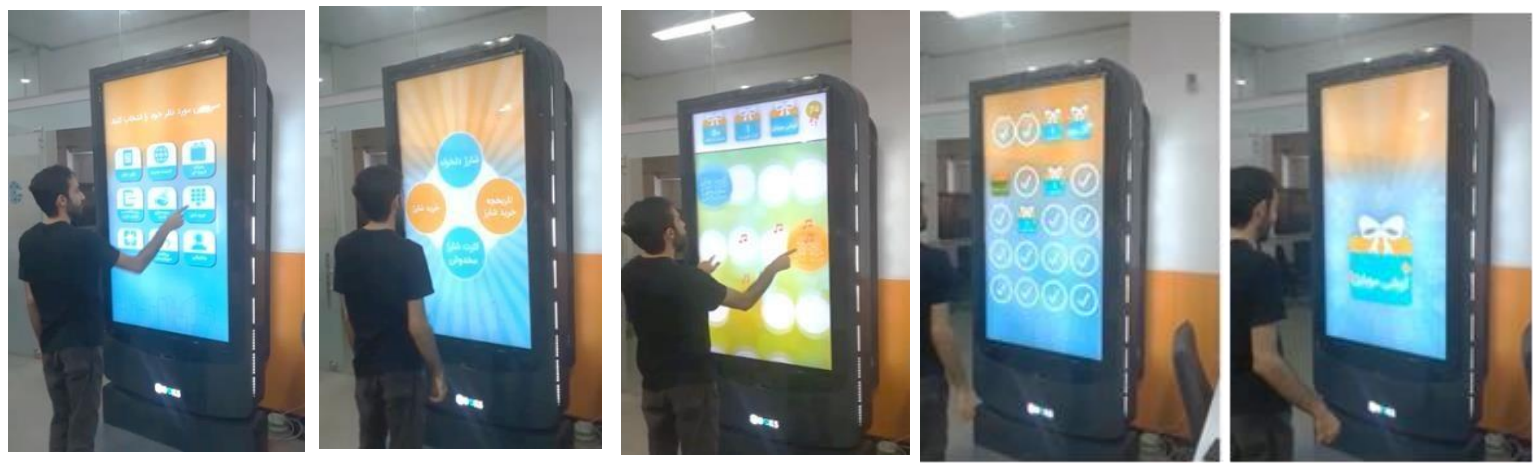

*Figure 2*. Left to right: activities and services related to the brand, the service provided features, the game, chance mechanism, the prize

## 3. Results

The results from a Chi-square test show that considering gender and age group, there is no significant difference between the two groups (p>0.05). According to product evaluation by individuals, satisfaction with advertising game, possibility of introducing to friends, and satisfaction with experience, there is a significant difference between two groups (advertising game with music and advertising game without music) (p<0.05). Descriptive results show that those who chose advertising game with music at 95% of confidence, have had a more positive evaluation of the product than those who chose advertising game without music. Also, satisfaction with advertising game in the first group (advertising game with music) at 95% of confidence, is higher than the second group (advertising game without music). Possibility of introducing to friends in the first group (advertising game with music) at 95% confidence is higher, comparing to the second group (advertising games without music). Also, satisfaction with experience in the first group (advertising game with music) at 95% of confidence is higher than the second group (advertising games without music).

Table 2: *Descriptive (Frequency and Percentage) and inferential Chi-Square Test results*

| Variable | Group | Advertising Game with music | | Advertising Game without music | | Chi-square Test results |
|---|---|---|---|---|---|---|
| | | Frequency | Percentage | Frequency | Percentage | |
| Product evaluation by individuals | Very Satisfied | 53 | 42.7 | 8 | 11.0 | $X^2=49.274$, P=0.001<0.05 |
| | Very Dissatisfied | 3 | 2.4 | 9 | 12.3 | |
| | Satisfied | 39 | 31.5 | 13 | 17.8 | |
| | Dissatisfied | 9 | 7.3 | 20 | 27.4 | |
| | Neutral | 20 | 16.1 | 23 | 31.5 | |
| | Total | 124 | 100.0 | 73 | 100.0 | |
| Satisfaction with advertising games | Yes | 106 | 85.5 | 27 | 37.0 | $X^2=49.274$, P=0.000<0.05 |
| | No | 18 | 14.5 | 46 | 63.0 | |
| | Total | 124 | 100.0 | 73 | 100.0 | |
| Possibility of introducing to | Not at all | 3 | 2.4 | 12 | 16.4 | $X^2=65.809$, P=0.000<0.05 |
| | <10 | 4 | 3.2 | 6 | 8.2 | |
| friends | 10_20 | 2 | 1.6 | 11 | 15.1 | |
| | 20_30 | 6 | 4.8 | 13 | 17.8 | |
| | 30_40 | 5 | 4.0 | 9 | 12.3 | |
| | 40_50 | 7 | 5.6 | 3 | 4.1 | |
| | 50_60 | 7 | 5.6 | 5 | 6.8 | |
| | 60_70 | 13 | 10.5 | 4 | 5.5 | |
| | 70_80 | 18 | 14.5 | 4 | 5.5 | |
| | 80_90 | 33 | 26.6 | 3 | 4.1 | |
| | >90 | 26 | 21.0 | 3 | 4.1 | |
| | Total | 124 | 100.0 | 73 | 100.0 | |
| Satisfaction with experience | Good | 36 | 29.0 | 16 | 21.9 | $X^2=56.438$, $P=0.000<0.05$ |
| | Poor | 5 | 4.0 | 22 | 30.1 | |
| | Excellent | 60 | 48.4 | 5 | 6.8 | |
| | Average | 23 | 18.5 | 30 | 41.1 | |
| | Total | 124 | 100.0 | 73 | 100.0 | |
| How much do you remember the ad? | You don't remember at all. | 5 | 4.0 | 19 | 26.0 | $X^2=48.86$, $P=0.000<0.05$ |
| | You remember the company but no product or advertisement | 8 | 6.5 | 14 | 19.2 | |
| | You remember the company and product but not advertising game | 20 | 16.1 | 22 | 30.1 | |
| | You remember the advertising game | 91 | 73.4 | 18 | 24.7 | |
| | Total | 124 | 100.0 | 73 | 100.0 | |

After 10 days, the participants were asked again if they remembered the game. The difference between the two groups is statistically significant (p<0.05). The result shows that the proportion of not remembering variable in the first group (advertising game with music) is just 4 percent, while the value of this variable in the second group (advertising games without music) is 26%. 6.5% of individuals in the first group remembered the company but not product or advertisement, while this value was 19.2% for the second group. 16.1% of the first group and 30.1% of the second group remembered the company and product but not the advertising game. Finally, 73.4% of the first group remembered the advertising game, while this value was 24.7% for the second group.

Our results also showed that the audience were more likely to play the Piano tiles with background music than without.

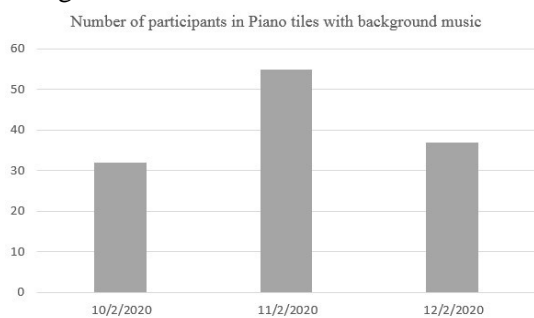

| Time | Participants |
|---|---|
| 10/2/2020 | 32 |
| 11/2/2020 | 55 |
| 12/2/2020 | 37 |
| Total | 124 |

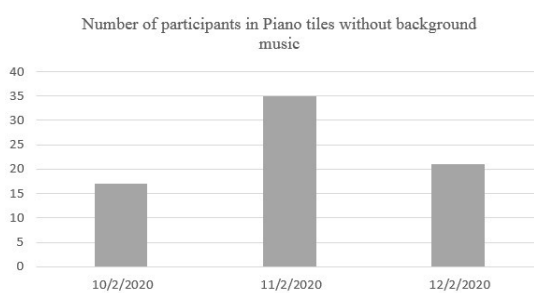

| Time | Participants |
|---|---|
| 10/2/2020 | 17 |
| 11/2/2020 | 35 |
| 12/2/2020 | 21 |
| Total | 73 |

*Figure 3a & 3b.* Number of participants in Piano tiles with and without background music over three days

## 4. Discussions & Future Work

We conducted a study to examine the uptake of a game with or without background music, in terms of the number of participants in an advergame. Our findings show that music in advergame is effective in attracting the audience and enhancing their user experience. Going forward, we will add personalisation to the games to see if that makes them more persuasive in engaging customers and enhancing their education of the brand awareness.

## Acknowledgements

We would like to thank Boks corp. for sharing their data and Dr Atefeh Ahmadi for her contribution to this project.